# A sequential surrogate method for reliability analysis based on radial basis function

Xu Li[1], Chunlin Gong[1], Liangxian Gu[1], Wenkun Gao[1], Zhao Jing[2], Hua Su[1]

[1]School of Astronautics, [2]School of Aeronautics

Northwestern Polytechnical University, Xi'an, P.R. China

**Abstract**: A radial basis function (RBF) based sequential surrogate reliability method (SSRM) is proposed, in which a special optimization problem is solved to update the surrogate model of the limit state function (LSF) iteratively. The objective of the optimization problem is to find a new point to maximize the probability density function (PDF), subject to the constraints that the new point is on the approximated LSF and the minimum distance to the existing points is greater than or equal to the given distance. By updating the surrogate model with the new points, the surrogate model of the LSF becomes more and more accurate in the important region with a high failure probability and on the LSF boundary. Moreover, the accuracy of the unimportant region is also improved within the iteration due to the minimum distance constraint. SSRM takes advantage of the information of PDF and LSF to capture the failure features, which decreases the number of the expensive LSF evaluations. Six numerical examples show that SSRM improves the accuracy of the surrogate model in the important region around the failure boundary with small number of samples and has better adaptability to the nonlinear LSF, hence increases the accuracy and efficiency of the reliability analysis.

## 1 Introduction

Reliability analysis is essentially a high-dimensional integration of complex and implicit limit state function (LSF). The numerical integration and Monte Carlo Simulation (MCS) with the original LSF face enormous computational challenges [1-2], therefore different approximations of the LSF are adopted to improve the computational efficiency.

The mean value method (MVM) [3-4] performs a first-order Taylor expansion of the LSF at the design point, in which the LSF is assumed to be normal distribution. MVM is one of the most efficient reliability methods. However, it requires the independent input variables obey normal distribution, which is difficult to be satisfied in practical problems. Moreover, the approximation error increases with the increase of the nonlinearity. Therefore, this method is suitable for fast estimation of the structural reliability with low nonlinearity.

---

*Corresponding author.
Email-mail address: lukebest@mail.nwpu.edu.cn

The first-order reliability method (FORM) [5-8] transforms random variables with different distributions into the same standard normal space **U** by Rosenblatt transformation [9] and then performs a first-order Taylor expansion at the most probable point (MPP) which has the maximum failure probability on the LSF. Eventually, the normal distribution parameters (mean value and standard deviation) of LSF are figured out with the gradient of the approximation function, and then the probability of failure is obtained. Compared with the MVM, the FORM doesn't require the input variables to obey normal distribution and has a higher accuracy with the LSF expanded at the MPP. However, the optimization with an equality constraint to find the MPP increases the number of the LSF evaluations. Moreover, it increases the nonlinearity of the LSF g(**x**) during the Rosenblatt transformation, thus the approximation error is large when the nonlinearly of LSF is high [10-11].

The second-order reliability method (SORM) [7, 12-13] is similar to the FORM. First, the input random variables are transformed into the standard normal space **U** to get the LSF G(**u**). Second, G(**u**) is expanded with second-order Taylor expansion at the MPP to obtain a quadratic hyper surface. Finally, the reliability of the approximate model is analyzed with analytical methods [10]. The SORM has a higher nonlinear adaptability than the FORM, but it needs to perform the time-consuming second-order gradients. Moreover, the adaptability to nonlinear boundary is still limited [10-11].

The response surface is another commonly used reliability analysis method [14-16]. By sampling in the neighborhood of the design point, the local approximation model of LSF is constructed. Since the response surface model evaluation time is far less than that of the original LSF, it is possible to use the approximate model for Mont Carlo Simulation (MCS) or Importance Sampling (IS) [17-19]. However, since the accuracy of the response surface method is poor with numbered samples, the complex structural behavior might not be captured. When considering the factors such as failure boundary and probability density, more samples are required to improve the accuracy. Therefore, the iterative response surface reliability analysis methods are proposed [20-30], which increases the samples in the important region near the MPP, and gradually improves the accuracy of the approximation model. This type of iterative method is also called adaptive method or active learning [31-32, 45]. In general, the response surface approximation model can be replaced by other surrogate model (also known as meto-model) techniques such as radial basis function (RBF), Kriging, support vector regression (SVR), artificial neutral net (ANN), etc. [18, 33-44]. The existing methods converge in a local region after increasing the sample density of the important region in some degree, but the accuracy doesn't increase in the less important region.

In order to make full use of the information of the added samples to improve the approximate accuracy of the important and less important region, this paper proposes a sequential surrogate reliability method (SSRM) based on RBF. By adding the points sequentially to the surrogate model, the failure features in the important region and on the boundary of the LSF are captured, and the failure probability is obtained with MCS by using the surrogate model. SSRM doesn't need to solve the MPP directly with the original LSF, but gradually approaches the important region near the MPP in the process of adding points, which reduces the number of sample evaluations and avoids the failure to find the optimal solution. Meanwhile, the SSRM method makes a trade-off between the precision and computational cost.

The remainder of this paper is structured as follows. **Section 2** introduces the surrogate model technology used in SSRM; **Section 3** describes the implementation process of SSRM, and analyzes its characteristics; **Section 4** verifies the effectiveness and efficiency of SSRM with six numerical examples; Conclusions are given in **Section 5**.

2 **Surrogate model**

2.1 Construction of surrogate model

The goal of the surrogate model is to construct a prediction model of a complex or unknown model with the observation samples. Instinctually, surrogate model is an interpolation or regression model, which is also a branch of machine learning [47]. The common surrogate models include polynomial response surface method (PRSM) [48-49], radial basis function (RBF) [48-49], Kriging [50-52], support vector regression (SVR) [50-52] and artificial neutral net (ANN) [53]. Since RBF has good nonlinear adaptability and is easy to implement, this paper constructs the sequential surrogate model with RBF. The observation samples are presented as

$$S = \{(\mathbf{x}_i, y_i) \mid i = 1, 2, \ldots, n\} \qquad (1)$$

denoted by the matrix form

$$\mathbf{X} = [\mathbf{x}_1, \mathbf{x}_2, \ldots, \mathbf{x}_n]^T$$
$$\mathbf{y} = [y_1, y_2, \ldots, y_n]^T \qquad (2)$$

where $n$ is the sample size; $\mathbf{X}$ is the input sample matrix; $\mathbf{y}$ is the output sample vector. $\mathbf{x}$ is a $m$-dimensional design variable, and the observed value at the point $\mathbf{x}$ is $\hat{y}(\mathbf{x})$. RBF uses a series of linear combinations of radial basis functions to approximate the expensive model. The basic expression is

$$\hat{y}(\mathbf{x}) = \sum_{i=1}^{n} \beta_i f(\|\mathbf{x} - \mathbf{x}_i\|) = \mathbf{f}(\mathbf{x})^T \boldsymbol{\beta} \qquad (3)$$

Where $\beta_i$ is the radial base coefficient; $f$ is the radial basis function, and the common forms are given in Table 1.

Table 1 Radial basis functions.

| Type | Function form |
| --- | --- |
| Gaussian | $e^{-cr^2}$ |
| Inverse Multiquadric | $\dfrac{1}{\sqrt{r^2 + c^2}}$ |
| Thin plate spline | $r^2 \log(cr^2 + 1)$ |

Substitute the samples of Eq. (1) into Eq. (3),

$$\begin{bmatrix} y_1 \\ y_2 \\ \vdots \\ y_n \end{bmatrix} = \begin{bmatrix} f(\|\mathbf{x}_1 - \mathbf{x}_1\|) & f(\|\mathbf{x}_1 - \mathbf{x}_2\|) & \cdots & f(\|\mathbf{x}_1 - \mathbf{x}_n\|) \\ f(\|\mathbf{x}_2 - \mathbf{x}_1\|) & f(\|\mathbf{x}_2 - \mathbf{x}_2\|) & \cdots & f(\|\mathbf{x}_2 - \mathbf{x}_n\|) \\ \vdots & \vdots & \ddots & \vdots \\ f(\|\mathbf{x}_n - \mathbf{x}_1\|) & f(\|\mathbf{x}_n - \mathbf{x}_2\|) & \cdots & f(\|\mathbf{x}_n - \mathbf{x}_n\|) \end{bmatrix} \begin{bmatrix} \beta_1 \\ \beta_2 \\ \vdots \\ \beta_n \end{bmatrix} \qquad (4)$$

denoted by the matrix form

$$\mathbf{y} = \mathbf{F}\boldsymbol{\beta} \qquad (5)$$

Since $\mathbf{F} \in R^{n \times n}$ is a non-singular matrix, Eq. (5) has a unique solution $\boldsymbol{\beta} = \mathbf{F}^{-1}\mathbf{y}$. Thus the prediction model is given by

$$\hat{y}(\mathbf{x}) = \mathbf{f}(\mathbf{x})^T \mathbf{F}^{-1} \mathbf{y} \qquad (6)$$

where $\mathbf{f}(\mathbf{x})$ is related to the prediction point $\mathbf{x}$ and sample input matrix $\mathbf{X}$; $\mathbf{F}^{-1}\mathbf{y}$ is only related to $\mathbf{X}$ and **y**. For a new prediction sample **x**, **f(x)** is calculated one time to get its predicted value $\hat{y}(\mathbf{x})$. It should be pointed out that the shape parameter *c*, which has a great influence on the accuracy of the model, is included in **F** and is determined by experience or other optimization criteria. This paper uses the cross-validation criteria to optimize the shape parameter *c*.

### 2.2 Validation of surrogate model

A common method to estimate the accuracy of the surrogate model is the root mean square errors (RMSE). However, since RBF requires the model strictly goes through sample points, RMSE is a constant zero, leads to the failure of the model estimation. To avoid the problem, the cross validation method is adopted. The samples are divided into *K* roughly equal-sized parts. For the *k*th part (*k* = 1, 2, ··, *K*), the model is fitted with the other *K*−1 parts of the samples, and calculates the prediction error of the fitted model when predicting the *k*th part of the samples [47]. Cross-validation can fully reflect the matching degree of the samples and the model. In particular, when *K* is equal to the sample size *n*, it is called leave-one-out cross validation error (LOOCV) [47]. Thus the RBF shape parameter *c* can be estimated by the following sub optimization problem:

$$\min_{c} \text{LOOCV}(c) = \sum_{i=1}^{n} \left[ y_i - \hat{y}\left(\mathbf{x}_i; S - \{(\mathbf{x}_i, y_i)\}\right) \right]^2 \quad (7)$$

In Eq. (7), it can be seen that the evaluation of LOOCV error requires *n* times construction of the surrogate model. However, the LOOCV error doesn't require additional verification points, which is capable of describing the match degree between the samples and the prediction model. According to the surrogate model and obtained parameter *c*, a reliability analysis method based on the sequential surrogate model is constructed.

### 3 Sequential surrogate reliability method

The SSRM uses the sequential surrogate model to increase the samples of the important region and the LSF boundary according to the specific optimization criteria, while reduces the samples in the unimportant region with small failure probability. In the process of iteration, the failure boundary of the LSF is approached step by step, and eventually an approximation LSF is obtained.

In order to facilitate the description of the algorithm, the original random variable $X_i$, according to Rosenblatt transformation [9], are transformed into standard normal distribution space **U**, where the independent random variables are $U_i \sim \mathbb{N}(0,1)$, $u_i = F_{U_i}^{-1} F_{X_i}(x_i), i = 1, 2, \ldots, n$. Therefore, the LSF is given by $g(\mathbf{x}) = g\left[F_\mathbf{X}^{-1} F_\mathbf{U}(\mathbf{u})\right] = G(\mathbf{u})$. The following algorithm is implemented in the **U**-space. The key step of SSRM is the iteration process with the special initial and terminal conditions, which can be briefly described as follows:

$$\begin{cases} \Delta \mathbf{u}_k = \underset{\mathbf{u}}{\operatorname{argmin}} \|\mathbf{u}\|, s.t. \begin{cases} \hat{G}(\mathbf{u}; S_k) = 0 \\ \underset{\mathbf{u}_i}{\min} \|\mathbf{u} - \mathbf{u}_i\| \geq d_{min} \\ \mathbf{u}_L \leq \mathbf{u} \leq \mathbf{u}_U \end{cases} \\ \Delta S_k = \{(\Delta \mathbf{u}_k, G(\Delta \mathbf{u}_k))\} \\ S_{k+1} = S_k \cup \Delta S_k \\ P_{k+1} = P_{\text{MCS}}\{\hat{G}(\mathbf{u}; S_{k+1}) \leq 0\} \\ \text{Start: } S_0, P_0 = P_{\text{MCS}}\{\hat{G}(\mathbf{u}; S_0) \leq 0\} \\ \text{Stop: } k \geq k_{\max} \text{ or } \left(|P_{k+1} - P_k| \leq \varepsilon_a \text{ and } \dfrac{|P_{k+1} - P_k|}{P_{k+1}} \leq \varepsilon_r\right) \end{cases} \tag{8}$$

Where $\Delta \mathbf{u}_k$ is the samples added in the $k$th iteration; $\|\mathbf{u}\| = \sqrt{\mathbf{u}^T \mathbf{u}}$ denotes the variable associated with probability density function (PDF) of the input random variables given by $\text{PDF}(\mathbf{u}) = \left(\dfrac{1}{\sqrt{2}}\right)^n \exp\left(-\dfrac{\|\mathbf{u}\|_2^2}{2}\right)$, therefore, PDF has the maximum value when $\|\mathbf{u}\|$ is minimum; $S_k$ is the sample set before the $k$th iteration; $\hat{G}(\mathbf{u}; S_k)$ represents the surrogate model of LSF constructed by $S_k$; $\mathbf{u}_i$ is the input variable of the sample set; $d_{min}$ is the minimum distance between the currently added point and the existing points; $P_{k+1}$ is the failure probability after the $k$th surrogate model update; $S_0$ is the initial sample set; $P_0$ is the reliability obtained from the reliability analysis based on the surrogate model established by the initial sample set; $k_{\max}$ is the required maximum number of points. The detailed steps of the algorithm are summarized as follows

**Step1**: Selecting the initial sample points by the Latin hypercube sampling (LHS) [46] and carrying out the evaluations of LSF to determine the initial sample set $S_0$. For the problem with small standard deviation, the initial sample number can be taken as $|S_0| = m+1$, while for the problem with large standard deviation, the initial sample number can be set as $|S_0| = 2m+1$, where $m$ is the dimension of the design variable;

**Step2**: Using the initial samples to construct surrogate model of the LSF $\hat{G}(\mathbf{u}; S_0)$, then the model shape parameter $c$ is optimized. Here surrogate model uses RBF which strictly goes through the sample points, and has strong nonlinear adaptive ability;

**Step3**: Using the initial surrogate model for MCS to obtain the initial failure probability $P_0 = P\{\hat{G}(\mathbf{u}; S_0) \leq 0\}$;

**Step4**: Solving the optimization problem to find the new point to be added. Since the problem has strong equality and inequality constraints which are not differentiable, the genetic algorithm (GA) is adopted. As the surrogate model is computationally cheap, the total optimization time is negligible compared with the direct LSF evaluation. With the execution of iteration, the subsequent increase of points can guarantee the accuracy of the important region. Therefore, although the optimization problem may be difficult to converge, the add points are not required to satisfy the constraints strictly. Furthermore, the minimum distance constraint ensures the algorithm not to fall into the local optimum.

**Step5**: Estimating the point found in **Step4** to get the added sample set $\Delta S_k = \{(\Delta \mathbf{u}_k, G(\Delta \mathbf{u}_k))\}$ and add it to the sample set $S_{k+1} = S_k \cup \Delta S_k$;

**Step6**: Updating the surrogate model of LSF $\hat{G}(\mathbf{u}; S_{k+1})$;

**Step7**: Estimating the failure reliability with the updated surrogate model by MCS, which can be presented as $P_{k+1} = P\{\hat{G}(\mathbf{u}; S_{k+1}) \leq 0\}$. In order to eliminate the errors induced by different samples, the MCS reliability analysis uses the same random seed, say, the same samples;

**Step8**: Convergence check. If one of the termination conditions, (a) the maximum iteration condition $k \geq k_{max}$, (b) the relative convergence condition $|P_{k+1} - P_k|/|P_{k+1}| \leq \varepsilon_r$ and the absolute convergence condition $|P_{k+1} - P_k| \leq \varepsilon_a$, is satisfied, go to **Step9**, otherwise set $k=k+1$ and turn to **Step4**;

**Step9**: End.

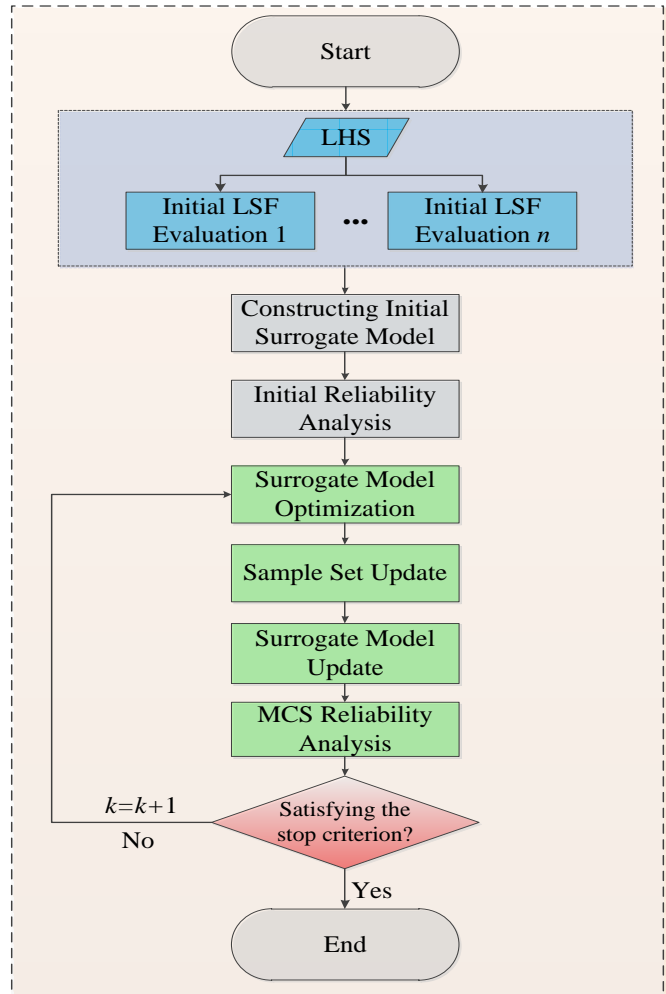

Fig. 1 Flow chart of SSRM.

The main procedures of SSRM are the construction of surrogate model and the criterion to add points, which are also the core to guarantee the accuracy and efficiency in the iteration process. SSRM automatically increases the samples in the important region with high failure probability, which makes full use of the information of PDF and each sample. In addition, since the minimum distance constraint of the samples works, SSRM increases the number of the samples in the less important region when the importance region includes enough points. Therefore, SSRM doesn't need many initial samples. After each surrogate model update, MCS reliability analysis is performed to achieve the estimation of failure probability. Since most of the failure features near the MPP and the LSF boundary are captured by the surrogate model iteratively, the MCS based estimation has a good accuracy. Moreover, since the time of a single evaluation of the surrogate model is far less than that of the original LSF, the cost of MCS based on the surrogate model is acceptable. Therefore, SSRM has a good accuracy and efficiency.

## 4 Numerical examples

In this section, six examples are used to perform SSRM, and the results are compared with those of MCS and some other existing methods. Example 1~3 are of different nonlinearity, example 4 is the LSF of a speed reducer shaft with variables under different distributions, example 5 is a cantilever tube structure with high-dimensional variable space and example 6 is a nonlinear oscillator with six random variables. Assuming the transformed random variables in **U**-space to be $u_i \in [-5,5] (i=1,2,\cdots,m)$, the initial samples are selected by LHS, and the sample size is $m+1$ or $2m+1$, where $m$ is the number of the random variables. In these examples, FORM and SORM are realized by Isight5.6, a software for multidisciplinary design optimization (MDO), while SSRM and MCS are implemented with the authors' in-house Matlab codes.

### 4.1 Circular pipe structure

Considering a circular pipe with circumferential through-wall crack subjected to a bending moment, the LSF is given by [4]:

$$g(X) = 4t\sigma_f R^2 \left( \cos\left(\frac{\theta}{2}\right) - \frac{1}{2}\sin(\theta) \right) - M \qquad (9)$$

where $\sigma_f$, $\theta$, $M$, $R$ and $t$ are flow stress, half-crack angle, applied bending moment, radius of the pipe and thickness of the pipe, respectively. $R$=0.3377m, $t$=0.03377m, $M$=3×10$^6$Nm. $\sigma_f$ and $\theta$ are random variables. The statistical properties of the random variables involved in the problem are depicted in Table 2.

Table 2 Distributions of random variables for the circular pipe structure.

| Variables | Mean | Standard deviation | Distribution |
|---|---|---|---|
| $\sigma_f$ | 301.079 | 14.78 | Normal |
| $\theta$ | 0.503 | 0.049 | Normal |

Table 3 Comparison of different reliability methods for the circular pipe structure.

| Methods | $P_f$ | Relative Error (%) | Number of LSF evaluations |
|---|---|---|---|
| FORM | 0.033065 | 3.7493 | 9 |
| SORM | 0.034211 | 0.4134 | 14 |
| SSRM | 0.034347 | 0.0175 | 7 |
| MCS | 0.034353 | 0.000 | 1×10$^6$ |

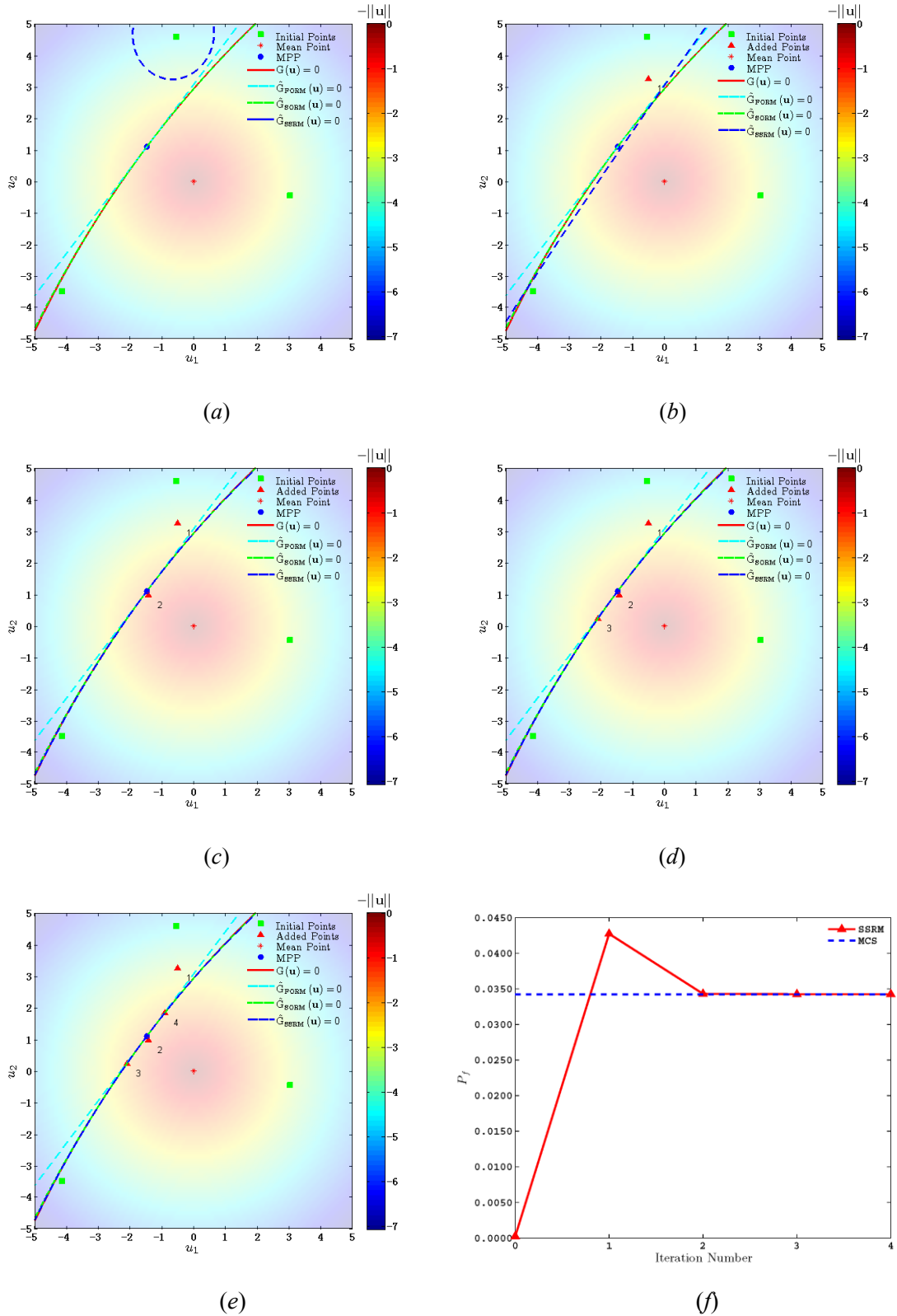

(*a*)                                        (*b*)

(*c*)                                        (*d*)

(*e*)                                        (*f*)

Fig. 2 Iterative process of SSRM for the circular pipe structure.

(*a*)~(*e*) Samples and LSF in **U**-space; (*f*) Iterative process of $P_f$

As shown in Fig. 2, 5 initial samples are selected and 4 extra samples are added iteratively. The approximate LSFs of FORM, SORM and SSRM are close to each other in the neighborhood of the MPP due to the low nonlinearity. In the region far away from the MPP, the approximate LSF of FORM has a poor accuracy, while SORM and SSRM have better accuracies with the approximate LSFs closer to the original LSF. As shown in Table 3, SSRM has the best accuracy with a relative error of 0.0175%. As the initial sample size is small, the initial surrogate LSF $G_{SSRM}(\mathbf{u})$ in Fig. 2 (a) is far away from the original LSF, however, after 4 iterations, they are almost the same.

### 4.2 Hyper-sphere bound problem

This examples considers a hyper-sphere bound with higher nonlinearity, and the LSF is given by

$$g(X) = 1 - X_1^3 - X_2^3 \tag{10}$$

Table 4 Distributions of random variables for the hyper-sphere problem.

| Variables | Mean | Standard deviation | Distribution |
|---|---|---|---|
| $X_1$ | 0.5 | 0.2 | Normal |
| $X_2$ | 0.5 | 0.2 | Normal |

Table 5 Comparison of different reliability methods for the hyper-sphere problem.

| Methods | $P_f$ | Relative Error (%) | Number of LSF evaluations |
|---|---|---|---|
| FORM | $1.891 \times 10^{-2}$ | 44.070 | 15 |
| SORM | $2.672 \times 10^{-2}$ | 20.970 | 20 |
| SSRM | $3.381 \times 10^{-2}$ | 0.000 | 12 |
| MCS | $3.381 \times 10^{-2}$ | 0.000 | $1 \times 10^6$ |

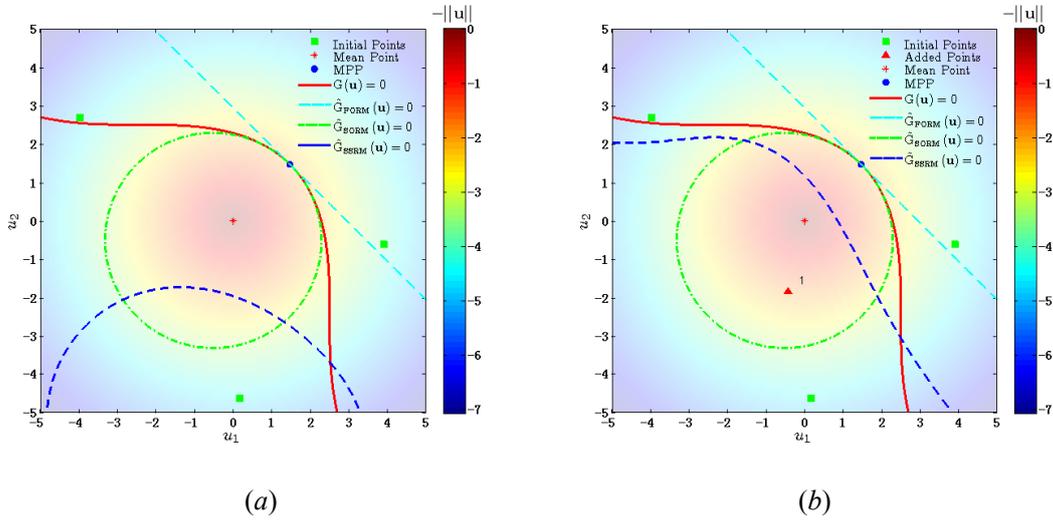

(a)  (b)

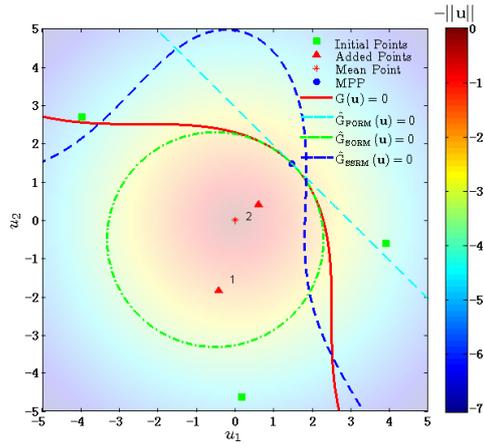

(c)

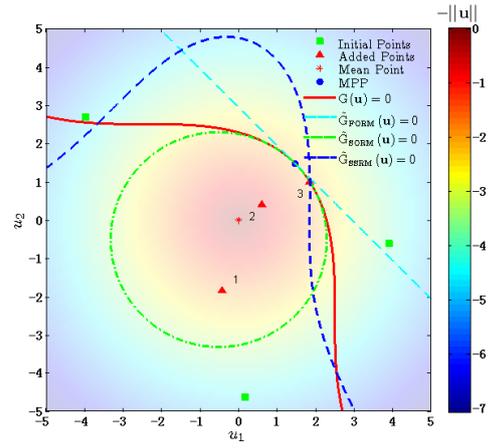

(d)

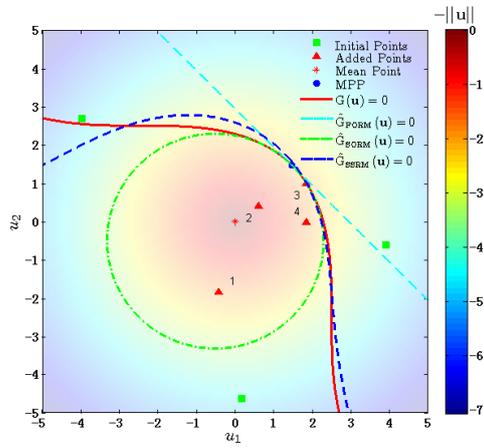

(e)

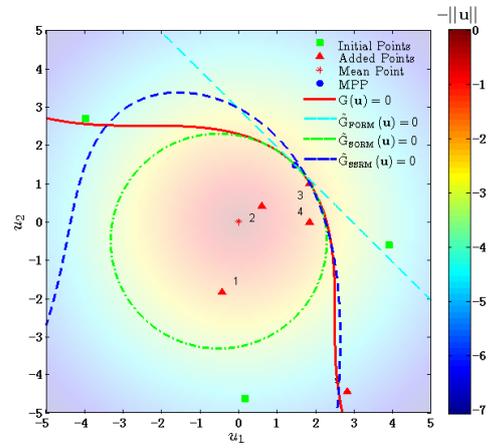

(f)

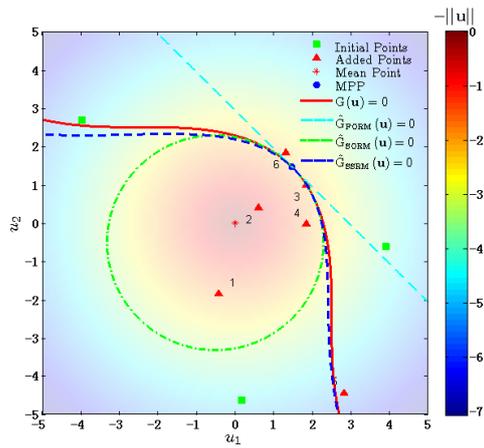

(g)

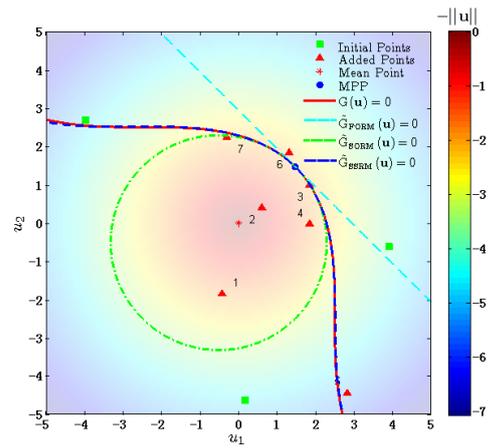

(h)

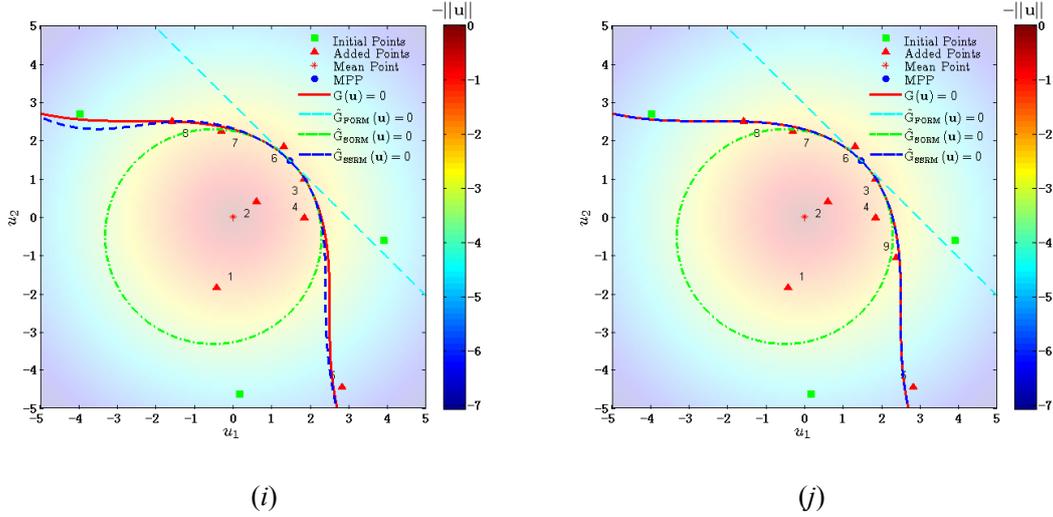

(*i*)            (*j*)

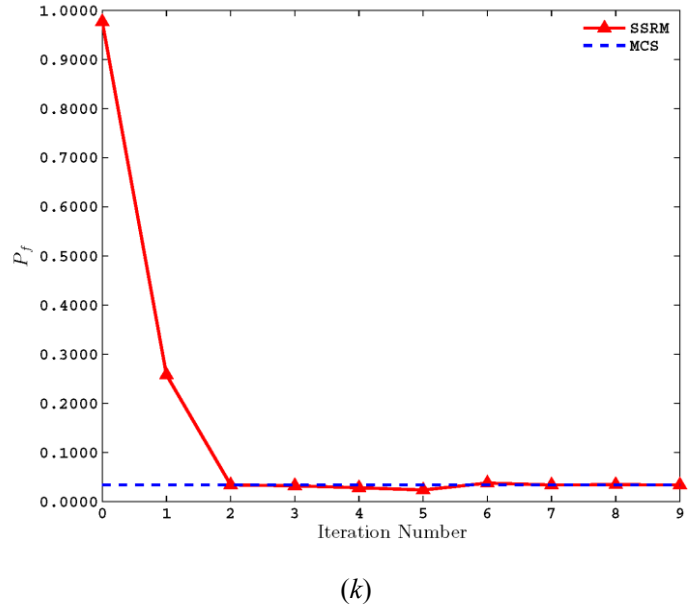

(*k*)

Fig. 3 Iterative process of SSRM for the hyper-sphere problem.

(*a*) ~ (*j*) Samples and LSF in **U**-space; (*k*) Iterative process of $P_f$.

As shown in Fig. 3, 3 initial samples are selected and 9 extra samples are added iteratively. As expected, Table 5 illustrates that SSRM obtains more accurate result with fewer samples. Since the approximation of the LSF can be improved step by step in the process of adding points, the accuracy is greatly improved. FORM and SORM cannot capture most of the failure features of the important region due to the high nonlinearity, however SSRM updates the surrogate of LSF, thus a good approximation in the important region and its neighborhood are obtained. As shown in Fig. 3 (a), the initial surrogate model is not accurate, and the added points are far away from the original LSF at first, however, as the iteration going on the added points get closer to the LSF. Thus each point makes contributions to the improvement of the accuracy of the LSF boundary, especially in the important region. Meanwhile, the region far away from the MPP and the LSF boundary has few samples and the probability is very small, hence the region has little effect on the estimation. That is why SSRM improves the accuracy with fewer samples.

## 4.3 Cantilever beam

This is also an example with higher nonlinearity, a cantilever beam with rectangular cross section subjected to uniformly distributed loading [20, 53]. The LSF with respect to the maximum deflection at the free end being greater than $l/325$, is given by

$$g = \frac{l}{325} - \frac{wbl^4}{8EI} \quad (11)$$

where $I=bh^3/12$, where $w, b, l, E, I$ and $h$ are load per unit, width, span, modulus of elasticity, moment of inertia of the cross section and depth, respectively. The random variables are $w$ and $h$, as shown in Table 6. Assuming $E$ and $l$ are $2.6\times10^4$ MPa and 6m respectively, the LSF is reduced to

$$g(x_1, x_2) = 18.46154 - 74769.23\frac{x_1}{x_2^3} \quad (12)$$

Table 6 Distributions of random variables for the cantilever beam.

| Variables | Mean | Standard deviation | Distribution |
|---|---|---|---|
| $x_1$ | 1000 | 200 | Normal |
| $x_2$ | 250.0 | 37.5 | Normal |

Table 7 Comparison of different reliability methods for the cantilever beam.

| Methods | $P_f$ | Relative Error (%) | Number of LSF evaluations |
|---|---|---|---|
| FORM | 0.00988 | 2.981 | 27 |
| SORM | 0.00988 | 2.981 | 32 |
| DS[a] | 0.01000 | 4.232 | 551 |
| DS+RS[b] | 0.00600 | 37.46 | 60 |
| DS+SP[c] | 0.00700 | 27.04 | 57 |
| DS+NN[d] | 0.00800 | 16.61 | 40 |
| IS[e] | 0.01000 | 4.232 | 9312 |
| IS+RS | 0.00900 | 6.191 | 2192 |
| IS+SP | 0.01000 | 4.232 | 358 |
| IS+NN | 0.01200 | 25.08 | 63 |
| SSRM | 0.009499 | 0.9902 | 18 |
| MCS | 0.009594 | 0.0000 | $1\times10^6$ |

[a]Directional Sampling; [b]Response Surface; [c]Splines; [d]Neural networks; [e]Important Sampling. These results are from Schueremans (2005)[53]

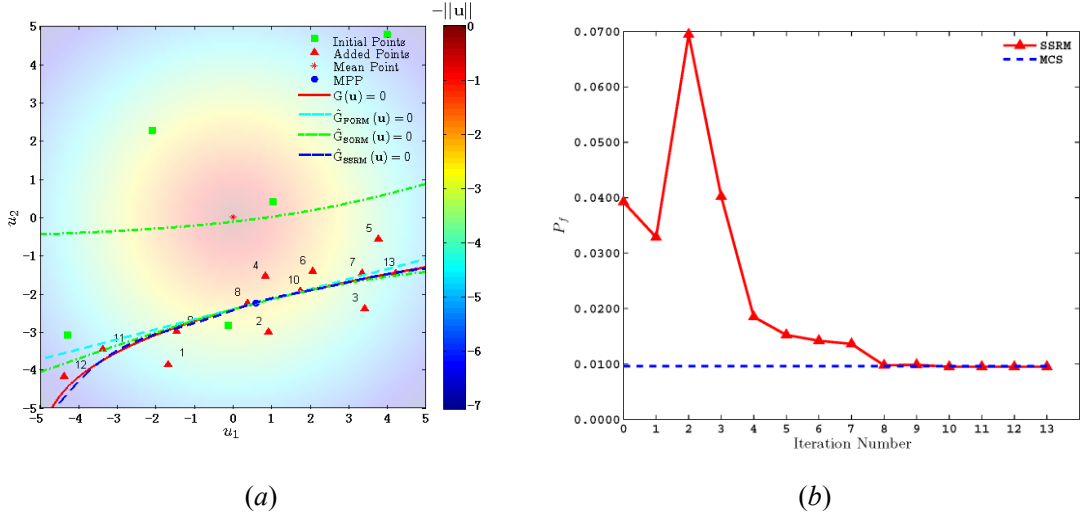

(*a*)                          (*b*)

Fig. 4 Iterative process of SSRM for the cantilever beam.
(*a*) Samples and LSF in **U**-space; (*b*) Iterative process of $P_f$.

As shown in Fig. 4, 5 initial samples are selected and 13 extra samples are added iteratively. Table 7 and Fig. 4 show that SSRM obtains a good accuracy with $P_f$=0.009499 and 18 LSF evaluations. Meanwhile, Fig. 4 (a) shows that the added samples are all near the LSF boundary, and keeps minimum distance to each other. It means that all the samples contribute a lot to obtain the failure probability with a good approximation to the LSF boundary. Moreover, it can be seen that the accuracy in the region near **u**= (-4, -4) is relatively poor. However, since the region is far from the mean point **u**= (0, 0), the probability is very small, which means that the poor accuracy doesn't affect the estimation of the failure probability.

4.4 Speed reducer shaft

In this section, considering the LSF of the reducer axis of multivariate random variables with different distributions [9]

$$g(X) = S - 32/(\pi D^3)\sqrt{F^2 L^2 /16 + T^2} \qquad (13)$$

Table 8 Distributions of random variables for the speed reducer shaft.

| Variables | Mean[a] \Low Bound[b] | Standard deviation[a] \Up Bound[b] | Distribution |
|---|---|---|---|
| Diameter $D$(mm) | 39 | 0.1 | Normal[a] |
| Span $L$(mm) | 400 | 0.1 | Normal[a] |
| External Force $F$(N) | 1500 | 350 | Grumble(Max)[a] |
| Torques $T$(Nmm) | 250000 | 35000 | Normal[a] |
| Strength $S$(MPa) | 70 | 80 | Uniform[b] |

[a]The distribution parameters are mean and standard deviation respectively.

Table 9 Comparison of different reliability methods for the speed reducer shaft.

| Methods | $P_f$ | Relative Error (%) | Number of LSF evaluations |
|---|---|---|---|
| FORM | 7.14392×10-4 | 90.50 | 30 |
| SORM | 7.12341×10-4 | 90.53 | 50 |
| SSRM | 7.52×10-3 | 0.00 | 44 |
| MCS | 7.52×10-3 | 0.00 | 1×106 |

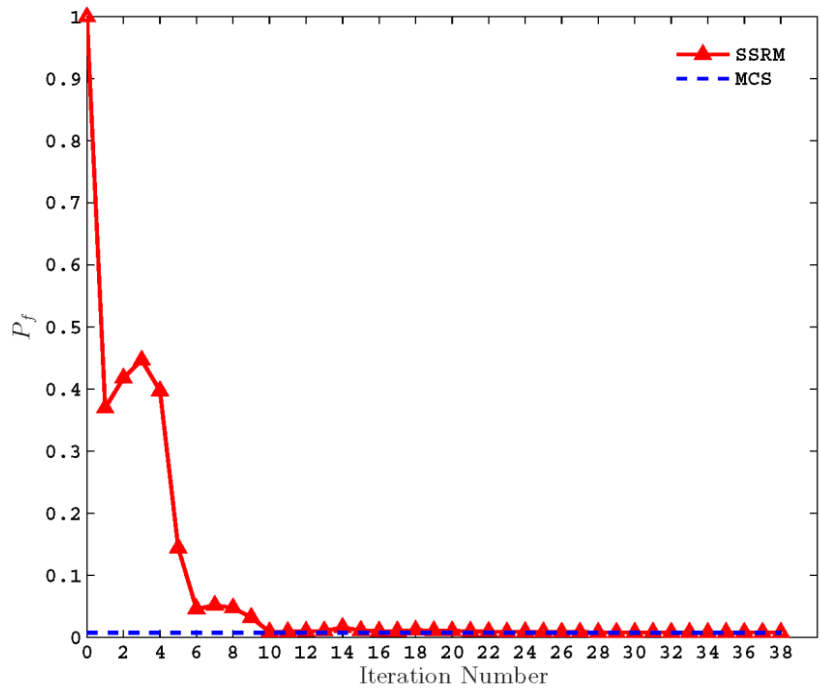

Fig. 5 Iterative process of SSRM for the speed reducer shaft.

In this example, 6 initial samples are selected and 38 extra samples are added iteratively. Fig. 5 shows that the initial value of $P_f$ is far from the MCS result, due to the small number of samples. However, with the increase of the samples during the iteration, the capture ability of the SSRM for the failure boundary is increase. Hence, $P_f$ converges in a few iterations. As expected, the results in Table 9 show that SSRM has a good adaptability and accuracy for this multidimensional problem.

4.5  Cantilever tube problem

To further demonstrate the performance of SSRM in the case of different distributions in high dimensions, consider the following cantilever tube problem, the LSF is defined as follows [6]:

$$g(X) = S_y - \sqrt{\sigma_x^2 + 3\tau_{zx}^2} \tag{14}$$

where

$$\begin{aligned}
&I = \frac{\pi}{64}\left[d^4 - (d-2t)^4\right] \\
&J = 2I \\
&\tau_{zx} = \frac{Td}{2J} \quad (15)\\
&M = F_1 L_1 \cos(\theta_1) + F_2 L_2 \cos(\theta_2) \\
&A = \frac{\pi}{4}\left[d^2 - (d-2t)^2\right] \\
&h = \frac{d}{2} \\
&\sigma_x = \frac{P + F_1 \sin(\theta_1) + F_2 \sin(\theta_1)}{A} + \frac{Mh}{I}
\end{aligned}$$

Constants $\theta_1 = 5°, \theta_2 = 10°$.

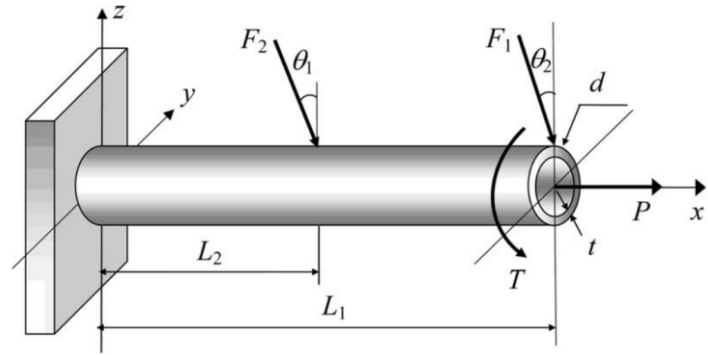

Fig. 6 Cantilever tube [6].

Table 10 Distributions of random variables for the cantilever tube.

| Variables | Mean[a] \Low Bound[b] | Standard deviation[a] \Up Bound[b] | Distribution |
|---|---|---|---|
| $t$(mm) | 5 | 0.1 | Normal[a] |
| $d$(mm) | 42 | 0.5 | Normal[a] |
| $L_1$(mm) | 119.75 | 120.25 | Uniform[b] |
| $L_2$(mm) | 59.75 | 60.25 | Uniform[b] |
| $F_1$(N) | 3000 | 300 | Normal[a] |
| $F_2$(N) | 3000 | 300 | Normal[a] |
| $P$(N) | 12000 | 1200 | Gumbel[a] |
| $T$(Nmm) | 90000 | 9000 | Normal[a] |
| $S_y$(MPa) | 220 | 22 | Normal[a] |

Table 11 Comparison of different reliability methods for the cantilever tube.

| Methods | $P_f$ | Relative Error (%) | Number of LSF evaluations |
|---|---|---|---|
| FORM | $3.8644 \times 10^{-4}$ | 63.06 | 30 |
| SORM | $3.8154 \times 10^{-4}$ | 63.52 | 84 |
| SSRM | $1.0520 \times 10^{-3}$ | 0.574 | 18 |
| MCS | $1.0460 \times 10^{-3}$ | 0.000 | $1 \times 10^6$ |

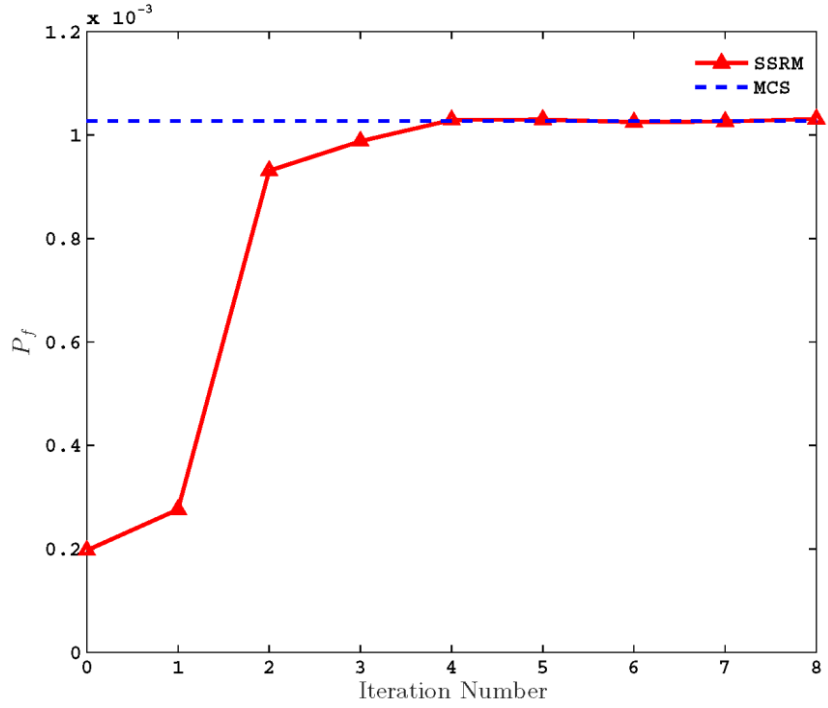

Fig. 7 Iterative process of SSRM for the cantilever tube.

In this example, 10 initial samples are selected and 8 extra samples are added iteratively. As a nine-dimensional problem, the computational cost for FORM and SORM to solve the MPP increases dramatically but not improve the accuracy effectively, while SSRM converges to the MCS result with 8 iterations (See Fig. 7). Though the dimension of the variables is high, the uniform distribution interval and the coefficient of variation are small, thus the nonlinearity near the design point is not very high. Therefore, the failure features can be captured by SSRM with a few samples; accordingly, a good estimation of the failure probability (See Table 11) is achieved.

4.6  Dynamic response of a nonlinear oscillator

This example consists of a nonlinear undamped single degree of freedom system with six random variables [20, 34, 53]. The LSF is given by

$$g(c_1, c_2, m, r, t_1, F_1) = 3r - \left| \frac{2F_1}{m\omega_0^2} \sin\left(\frac{\omega_0 t_1}{2}\right) \right| \quad (16)$$

where $\omega_0 = \sqrt{(c_1+c_2)/m}$. The random variables are shown in Table 12.

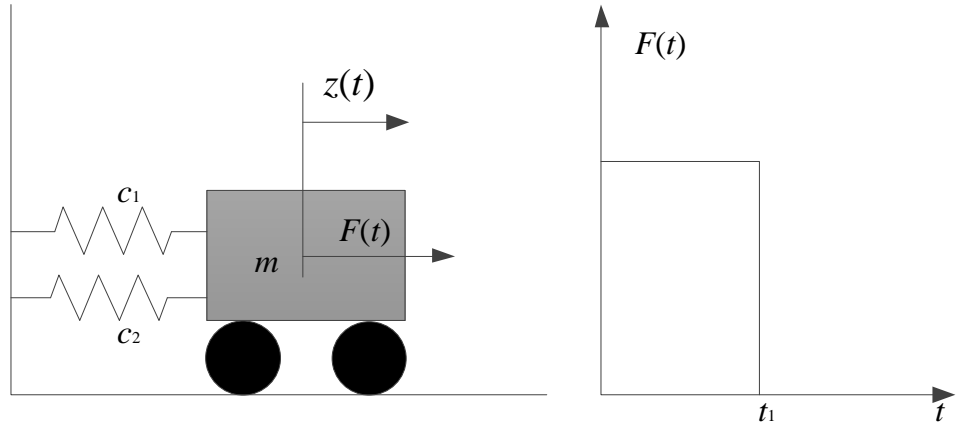

Fig. 8 Nonlinear oscillator [34].

Table 12 Distributions of random variables for the nonlinear oscillator.

| Variables | Mean | Standard deviation | Distribution |
|---|---|---|---|
| $m$ | 1 | 0.05 | Normal |
| $c_1$ | 1 | 0.1 | Normal |
| $c_2$ | 0.1 | 0.01 | Normal |
| $r$ | 0.5 | 0.05 | Normal |
| $F_1$ | 0.5 | 0.2 | Normal |
| $t_1$ | 1 | 0.2 | Normal |

Table 13 Comparison of different reliability methods for the nonlinear oscillator.

| Methods | $P_f$ | Relative Error (%) | Number of LSF evaluations |
|---|---|---|---|
| FORM | $3.108 \times 10^{-2}$ | 9.668 | 35 |
| SORM | $2.840 \times 10^{-2}$ | 0.212 | 48 |
| DS[a] | $3.500 \times 10^{-2}$ | 23.500 | 1281 |
| DS+RS[b] | $3.400 \times 10^{-2}$ | 19.972 | 62 |
| DS+SP[c] | $3.400 \times 10^{-2}$ | 19.972 | 76 |
| DS+NN[d] | $2.800 \times 10^{-2}$ | 1.200 | 86 |
| IS[e] | $2.700 \times 10^{-2}$ | 4.728 | 6144 |
| IS+RS | $2.500 \times 10^{-2}$ | 11.785 | 109 |
| IS+SP | $2.700 \times 10^{-2}$ | 4.728 | 67 |
| IS+NN | $3.100 \times 10^{-2}$ | 9.386 | 68 |
| AK[f]+MCS+U[g] | $2.834 \times 10^{-2}$ | 0.000 | 58 |
| AK+MCS+EFF[h] | $2.851 \times 10^{-2}$ | 0.600 | 45 |
| SSRM | $2.880 \times 10^{-2}$ | 1.623 | 19 |
| MCS(2.2%)[i] | $2.834 \times 10^{-2}$ | 0.000 | $7 \times 10^4$ |

[a]Directional Sampling; [b]Response Surface; [c]Splines; [d]Neural networks; [e]Important Sampling. These results are from Schueremans et. al. (2005) [53].

[f]Active Learning; [g]Learning Function U; [h]Expected Feasibility Function; [i]MCS with a covariance of 2.2%. These results are from Echard et. al. (2011) [34].

Table 14 Iteration process of SSRM for the nonlinear oscillator.

| Iteration No. | $u_1(m)$ | $u_2(c_1)$ | $u_3(c_2)$ | $u_4(r)$ | $u_5(F_1)$ | $u_6(t_1)$ | $P_f$ |
|---|---|---|---|---|---|---|---|
| 0 | 2.140 | -0.125 | -2.975 | 3.075 | -2.505 | -1.525 | |
| | -4.899 | 1.441 | -2.518 | 4.468 | 0.549 | 2.988 | |
| | 3.734 | 1.061 | 4.171 | 4.113 | 0.357 | 1.517 | |
| | 3.031 | -3.786 | 2.319 | -3.929 | -4.085 | -0.482 | |
| | -2.600 | 3.502 | 1.669 | -4.875 | 1.794 | 2.596 | |
| | -0.328 | 2.584 | -0.878 | 0.055 | 3.379 | -4.064 | |
| | -1.125 | -3.443 | -0.331 | -1.238 | -4.870 | 0.467 | 0.04303 |
| | -3.605 | 3.345 | -4.844 | 1.590 | -2.749 | -4.786 | |
| | 1.003 | -2.119 | -1.447 | -0.900 | 2.168 | 3.782 | |
| | -1.717 | -4.542 | 4.954 | -2.308 | -1.326 | -2.092 | |
| | 1.776 | -0.992 | 2.996 | 0.777 | 3.906 | 4.729 | |
| | -3.119 | -1.315 | 0.628 | 1.988 | -0.608 | 0.191 | |
| | 4.862 | 4.384 | -4.017 | -3.219 | 4.269 | -2.870 | |
| 1 | 0.109 | -0.428 | 0.152 | -0.161 | 1.307 | 1.136 | 0.02921 |
| 2 | 0.007 | -0.037 | 0.041 | -0.702 | 0.806 | 1.641 | 0.02789 |
| 3 | 0.325 | 0.761 | 0.012 | -0.008 | 1.546 | 1.947 | 0.02680 |
| 4 | -0.010 | -0.238 | 0.018 | -0.750 | 1.742 | 0.759 | 0.02904 |
| 5 | -0.074 | -1.080 | 0.016 | 0.254 | 1.600 | 1.105 | 0.02881 |
| 6 | 0.081 | -0.433 | -0.270 | -0.817 | 1.157 | 1.154 | 0.02880 |

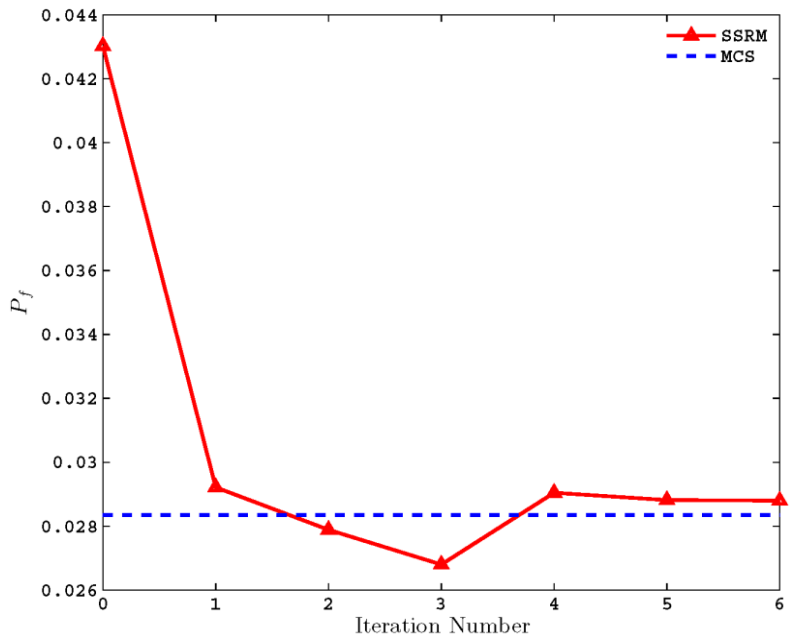

Fig. 9 Iterative process of SSRM for the nonlinear oscillator.

In this problem, 13 initial samples are selected and 6 extra samples are added as shown in Fig. 9. As expected, the $P_f$ of SSRM in Table 13 is also close to that of MCS with a relative error of 1.623%. Since the result of MCS is also with a variance of 2.2%, the relative error of SSRM is acceptable. However, the total LSF evaluations of SSRM are 19, which is smaller than the existing methods.

## 5 Conclusions

In this paper, an efficient reliability analysis method named SSRM based on RBF is proposed, which uses the probability density function and the sample density information to update the accuracy of the important and less important region by the sequential surrogate model. The surrogate model captures the failure characteristics in the important region, while reduces the samples in the region with low failure probability. Multiple numerical results demonstrate that SSRM has a good adaptability to the LSFs with different nonlinearity and variable dimensions. In general, the number of samples for SSRM to convergence increases with the nonlinearity of the model and the variable dimension. Moreover, due to the characteristics of the surrogate model, the increase of samples can be achieved in parallel, which means several points can be added at the same time to further enhance the convergence efficiency. When considering the system reliability, the optimum criterion to add the points is worth to explore, and further study should be focused on how to reduce the high-fidelity samples in variable-fidelity surrogate model.

## Acknowledgement

The authors gratefully appreciate the support by NPU Foundation for Fundamental Research [Grant No. G2016KY0302], and also thank Dr. Hua Su and Dr. Xueyu Li for the helpful discussion and constructive advice.